\pdfoutput=1
\documentclass[11pt,twoside,a4paper,cmspaper,final,collab]{cms-tdr}
\def\svnVersion{2a8f98b}\def\svnDate{2023/05/23}\def\cmsCernNoTag{CERN-EP-2022-088}\def\cmsCernDate{\today}\def\cmsMessage{Published in Physics Letters B as \href{http://dx.doi.org/10.1016/j.physletb.2022.137563}{\doi{10.1016/j.physletb.2022.137563}.}}
\begin{document}\cmsNoteHeader{SMP-18-014}

\newlength\cmsTabSkip\setlength{\cmsTabSkip}{1ex}
\newlength\cmsFigWidth
\ifthenelse{\boolean{cms@external}}{\setlength{\cmsFigWidth}{\columnwidth}}{\setlength{\cmsFigWidth}{0.8\textwidth}}
\newcommand{\zvvjets}{\ensuremath{\PZ(\PGn\PGn)\!+\!\text{jets}}\xspace}
\newcommand{\gstar}{\ensuremath{\gamma\mbox{*}}\xspace}
\newcommand{\dymmjets}{\ensuremath{\PZ/\gstar(\mu\mu)\!+\!\text{jets}}\xspace}
\newcommand{\dyeejets}{\ensuremath{\PZ/\gstar(\Pe\Pe)\!+\!\text{jets}}\xspace}
\newcommand{\ptmissjets}{\ensuremath{p_\text{T}^{\text{miss}}\!+\!\text{jets}}\xspace}
\newcommand{\zvv}{\ensuremath{\PZ \to \PGn\PAGn}\xspace}
\newcommand{\widthinv}{\ensuremath{\Gamma_{\text{inv}}}\xspace}
\newcommand{\widthinvMC}{\ensuremath{\Gamma^{\text{MC}}(\zvv)}\xspace}
\newcommand{\htmiss}{\ensuremath{H_\text{T}^{\text{miss}}}\xspace}
\newcommand{\ptmisscalo}{\ensuremath{p_\text{T,Calo.}^{\text{miss}}}\xspace}
\newcommand{\tw}{\ensuremath{t_{\PW}}\xspace}
\newcommand{\recoil}{\ensuremath{\mathcal{U}}\xspace}
\newcommand{\recoilvec}{\ensuremath{\vec{\mathcal{U}}}\xspace}
\newcommand{\dymm}{\ensuremath{\PZ/\gamma\mbox{*} \to \mu\mu}\xspace}
\newcommand{\dyee}{\ensuremath{\PZ/\gamma\mbox{*} \to \Pe\Pe}\xspace}
\newcommand{\mll}{\ensuremath{m_{\ell\ell}}\xspace}
\newcommand{\zjets}{\ensuremath{\PZ\!+\!\text{jets}}\xspace}
\newcommand{\zjet}{\ensuremath{\PZ\!+\!\text{jet}}\xspace}

\newcommand{\dyll}{\ensuremath{\PZ/\gamma\mbox{*} \to \ell\ell}\xspace}
\newcommand{\gstarll}{\ensuremath{\gamma\mbox{*} \to \ell\ell}\xspace}
\newcommand{\rinv}{\ensuremath{r_{\text{inv}}}\xspace}
\newcommand{\wjets}{\ensuremath{\PW\!+\!\text{jets}}\xspace}
\newcommand{\ejets}{\ensuremath{\Pe\!+\!\text{jets}}\xspace}
\newcommand{\mujets}{\ensuremath{\PGm\!+\!\text{jets}}\xspace}
\newcommand{\wlnujets}{\ensuremath{\PW (\ell\PGn)\!+\!\text{jets}}\xspace}
\newcommand{\tauhjets}{\ensuremath{\tauh\!+\!\text{jets}}\xspace}
\newcommand{\zll}{\ensuremath{\PZ \to \ell\ell}\xspace}
\newcommand{\branching}{\ensuremath{\mathcal{B}}\xspace}
\newcommand{\dphimin}{\ensuremath{\Delta\phi_{\text{min}}}}

\cmsNoteHeader{SMP-18-014}
\title{Precision measurement of the \texorpdfstring{\PZ}{Z} boson invisible width in \texorpdfstring{$\Pp\Pp$}{pp} collisions at \texorpdfstring{$\sqrt{s} = 13\TeV$}{sqrt(s) = 13 TeV}}

\date{\today}

\abstract{A precise measurement of the invisible width of the \PZ boson produced in proton-proton collisions at a center-of-mass energy of 13\TeV is presented using data recorded by the CMS experiment at the LHC, corresponding to an integrated luminosity of 36.3\fbinv. The result is obtained from a simultaneous fit to kinematic distributions for two data samples of \PZ boson plus jets: one dominated by \PZ boson decays to invisible particles and the other by \PZ boson decays to muon and electron pairs. The invisible width is measured to be $523\pm 3\stat\pm 16\syst\MeV$. This result is the first precise measurement of the invisible width of the \PZ boson at a hadron collider, and is the single most precise direct measurement to date, competitive with the combined result of the direct measurements from the LEP experiments.}

\hypersetup{%
pdfauthor={CMS Collaboration},%
pdftitle={Precision measurement of the Z invisible width with the CMS experiment in pp collisions at sqrt(s) = 13 TeV},%
pdfsubject={CMS},%
pdfkeywords={CMS, Z boson}
}

\maketitle

Precision measurements of \PZ boson properties at the CERN LHC enable important tests of the standard model (SM). Deviations from the SM predictions could reveal signs of new physics beyond the SM.
Among the physical observables describing the \PZ boson, the `invisible width' corresponds to \PZ boson decays to particles that are not detected, such as neutrino-antineutrino pairs, and can be translated into a constraint on the number of light neutrino species coupling to the \PZ boson. A precise measurement of this quantity could reveal non-SM contributions from new-physics scenarios~\cite{CMS:2021far}. 

Experiments at the CERN Large Electron-Positron Collider (LEP) measured the invisible width of the \PZ boson using both direct and indirect approaches. 
The direct method is based on the associated production of an initial-state photon with a \PZ boson that subsequently decays invisibly. Direct measurements have been made by the OPAL~\cite{Akers1995}, ALEPH~\cite{aleph1993520}, and L3~\cite{L3} experiments, yielding the LEP combined measurement of $503 \pm 16\MeV$~\cite{10.1093/ptep/ptaa104}. 
The indirect method uses the total \PZ boson width extracted from the \PZ boson lineshape, and subtracts the measured partial decay widths to all known visible final states.
This method is the most precise, with a combined indirect measurement of $499.0 \pm 1.5\MeV$~\cite{10.1093/ptep/ptaa104} from the LEP experiments. A previous measurement by the UA2 experiment used the \PW\ boson and \PZ boson widths to deduce a limit on the number of light neutrino generations~\cite{UA2:1987qqc}. 

This Letter presents the first direct measurement of the invisible width of the \PZ boson at a hadron collider.
The direct and indirect measurements could be sensitive to different new-physics scenarios~\cite{Carena:2003aj} motivating the goal to reduce the total uncertainty in the direct measurement.
We use data recorded by the CMS experiment from proton-proton ($\Pp\Pp$) collisions at a center-of-mass energy of 13\TeV, corresponding to an integrated luminosity of 36.3\fbinv~\cite{CMS:2021xjt}.
This measurement exploits the similarity in kinematic characteristics between the decay of the \PZ boson to neutrinos and its decay to charged leptons (in this case, electrons and muons), and is based on the ratio of branching fractions between these decay modes, as given by the following:
\begin{linenomath}
    \begin{equation}\label{eq:zinv-width}
        \Gamma(\zvv) = \frac{\sigma(\zjets) \branching(\zvv)}
        {\sigma(\zjets) \branching(\zll)}
        \Gamma(\zll)
    \end{equation}
\end{linenomath}

where $\sigma(\zjets)$ is the cross section to produce a \PZ boson in association with jets, $\Gamma(\zvv)$ and $\branching(\zvv)$ are the partial width and branching fraction of the \PZ boson to neutrinos. Similarly, for the charged leptons the partial width and branching fraction are $\Gamma(\zll)$ and $\branching(\zll)$. 
The invisible width, \widthinv, is extracted from a simultaneous fit to kinematic distributions for two data samples: one dominated by \PZ boson decays to invisible particles and the other by \PZ boson decays to muon and electron pairs. Since the invisible particles cannot be detected, invisible \PZ boson decays can only be identified when the \PZ boson has a significant transverse momentum (\pt) leading to large missing transverse momentum (\ptmiss). In this analysis, therefore, only \PZ bosons produced in association with jets are considered. 
Tabulated results are available in the HepData database~\cite{hepdata}.
A previous measurement by CMS of the differential production cross section of the invisible decay of the \PZ boson is presented in~\cite{CMS:2020hkd}.

The central feature of the CMS apparatus is a superconducting solenoid of 6\unit{m} internal diameter, providing a magnetic field of 3.8\unit{T}.
Within the solenoid volume are a silicon pixel and strip tracker, a lead tungstate crystal electromagnetic calorimeter (ECAL), and a brass and scintillator hadron calorimeter, each composed of a barrel and two endcap sections. Forward calorimeters extend the pseudorapidity ($\eta$) coverage provided by the barrel and endcap detectors.
Muons are detected in gas-ionization chambers embedded in the steel flux-return yoke outside the solenoid. Events of interest are selected using a two-tiered trigger system~\cite{Khachatryan:2016bia}.
A detailed description of the CMS detector and relevant kinematic variables is reported in Ref.~\cite{Chatrchyan:2008zzk}.

The particle-flow (PF) algorithm~\cite{CMS-PRF-14-001} reconstructs and identifies individual particles in an event with an optimized combination of information from the various elements of the CMS detector. It identifies these particles as charged or neutral hadrons, photons, and charged leptons. The energy of electrons is determined from a combination of the electron momentum measured by the tracker and the cluster of energy deposition in the ECAL. Muons are identified as tracks in the central tracker consistent with either a track or several hits in the muon system, and associated with energy deposits in the calorimeter compatible with the muon hypothesis. The momentum of muons is obtained from the curvature of the corresponding track. 
The primary vertex (PV) is taken to be the vertex corresponding to the hardest scattering in the event, evaluated using tracking information alone, as described in Section 9.4.1 of~\cite{Contardo:2020886}.

For each event, hadronic jets are clustered from these reconstructed particles using the infrared- and collinear-safe anti-\kt algorithm~\cite{Cacciari:2008gp, Cacciari:2011ma} with a distance parameter of 0.4. Jet energy corrections are derived from simulation studies so that the average measured response of jets match that of particle-level jets. In situ measurements of the momentum balance in dijet, photon+jet, \zjet, and multijet events are used to determine any residual differences between the jet energy scale (JES) in data and simulation, and appropriate corrections are made~\cite{Khachatryan:2016kdb}. The jet energy resolution (JER) typically amounts to 15\% at 10\GeV, 8\% at 100\GeV, and 4\% at 1\TeV.

Hadronic $\Pgt$ decays ($\tauh$) are reconstructed using the hadrons-plus-strips algorithm~\cite{CMS:2018jrd}, which combines 1 or 3 tracks with energy deposits in the calorimeters, to identify the tau decay modes. To distinguish genuine $\tauh$ decays from jets originating from the hadronization of quarks or gluons, and from electrons, a discriminant based on a multivariate analysis is used~\cite{CMS:2018jrd}. 

The vector \ptvecmiss is computed as the negative vector sum of the \pt of all particle candidates from the PF algorithm in an event, and its magnitude is denoted as \ptmiss~\cite{Sirunyan:2019kia}. The \ptvecmiss vector is modified to account for corrections to the energy scale of the reconstructed jets in the event. For the leptonic data sets selected in this analysis, \ptvecmiss is replaced by the recoil vector \recoilvec, defined as the vector sum of \ptvecmiss and \ptvec of the charged leptons passing the object selection in each event. This parameter acts as a proxy for the boson \pt. Henceforth $\recoil=\abs{\recoilvec}$ is used to denote \ptmiss for events with no leptons and \recoil for events containing leptons. 

The signal process of a \PZ boson decaying to neutrinos produced in association with jets is simulated at next-to-leading order (NLO) using the Monte Carlo (MC) event generator \MGvATNLO 2.3.2~\cite{Alwall:2014hca} interfaced with \PYTHIA (v8.212)~\cite{Sjostrand:2014zea} for parton shower (PS) and hadronization. The matrix element (ME) calculation includes up to three final-state partons generated from quantum chromodynamics (QCD) interactions and the ME-PS matching is performed following the FxFx~\cite{Frederix:2012ps} prescription.  
The Drell--Yan production of $\PZ/\gstar$ in association with jets, where the $\PZ/\gstar$ decays to charged leptons, and the production of a \PW\ boson in association with jets are simulated in the same way. 
Corrections are applied to the cross sections for these processes as a function of the boson \pt to account for higher-order effects from next-to-next-to-leading order (NNLO) in QCD and NLO in electroweak (EW)~\cite{Lindert:2017olm} interactions. The impact of the NLO EW corrections is particularly significant since they become large and negative at high \pt, leading to a 20--30\% correction for \zjets and \wjets production cross sections at 1\TeV in boson \pt~\cite{Lindert:2017olm}. 

Top quark pair events are generated with \MGvATNLO and normalized to the inclusive cross section calculated at NNLO matched to next-to-next-to-leading logarithmic accuracy~\cite{Kidonakis:2012db,Czakon:2013goa}. Single top quark processes are generated at leading order (LO) with \POWHEG~\cite{Nason:2004rx,Frixione:2007vw,Alioli:2010xd} and normalized to the NLO cross sections for $\PQt\PW$ and \PQt-channel production~\cite{Re:2010bp}, whereas the $s$-channel production is generated at NLO with \MGvATNLO.
The production of diboson processes is simulated using a combination of \MGvATNLO and \POWHEG at NLO accuracy. The photon+jets process is generated using the \MGvATNLO generator at NLO with up to one parton in the final state and the ME-PS matching is performed following the FxFx prescription.
QCD multijet production is generated with \PYTHIA at LO. 

The {NNPDF}3.0 LO and NLO~\cite{nnpdf} parton distribution functions (PDFs) are used, respectively, with the LO and NLO generators described above. The \PYTHIA program with the CUETP8M1 underlying event tune~\cite{Khachatryan:2015pea} is used to describe parton showering and hadronization for all simulated samples. The full detector response is simulated using the \GEANTfour~\cite{geant} package for all background and signal samples. 

Additional $\Pp\Pp$ interactions within the same or nearby bunch crossings (pileup) are included in the simulated samples, with the multiplicity weighted to match the
pileup distribution measured in data. 

The events used in this analysis can be categorized into three main regions: the \ptmiss plus jets region that is composed predominantly of \zvv events; a dilepton plus jets region with \dymm and \dyee events; and a single-lepton plus jets region, $\ell$+jets, where $\ell=\Pe, \PGm, \tauh$ (hadronically decaying $\tau$ lepton), that is enriched in \wjets events.

Events are selected using an unprescaled trigger that requires $\ptmiss>90\GeV$ and $\htmiss>90\GeV$, where \htmiss is the magnitude of the vectorial \pt sum of all jets with $\pt>20\GeV$ and $\abs{\eta}<5.0$. Muons are not included in the calculation of the \ptmiss and \htmiss variables at the trigger level. This enables the same trigger to be used to select events for the \ptmiss plus jets region and the \dymm and \mujets regions. This trigger is also used to select the \tauhjets region. 
Events for the \dyee and \ejets regions are selected using a trigger that requires at least one reconstructed electron passing a tight identification selection~\cite{CMS:2020uim} and with $\pt>27\GeV$.

Muon candidates are required to be reconstructed in the fiducial region of pseudorapidity $\abs{\eta}<2.4$ of the muon system.
Electron candidates are required to be reconstructed within $\abs{\eta}<2.4$, excluding the barrel-to-endcap $1.444<\abs{\eta}<1.566$ transition regions of the ECAL. 
The leptons are required to be isolated, where the isolation is calculated from PF candidates and corrected for pileup on an event-by-event basis. The scalar \pt sum over all particle candidates reconstructed in an isolation cone with radius $R = \sqrt{\smash[b]{(\Delta\eta)^2+(\Delta\phi)^2}} = 0.4$ (0.3), where $\phi$ is the azimuthal angle, around the momentum vector of the muon (electron) must be $< 15\% (6\%)$ of the lepton \pt. Jets are required to have $\pt>40\GeV$ and $\abs{\eta}<2.4$. 

A set of `baseline' selection criteria are applied to all regions. It requires \recoil to be $>200\GeV$, which ensures that the trigger efficiency is high ($ > 98\%$), and the leading jet in the event to have $\pt>200\GeV$. 
An event is vetoed if there are jets with $\pt>40\GeV$ and $\abs{\eta}>2.4$ to reduce the contribution from vector boson scattering processes. 
Misidentified high-\pt muons can produce \ptmiss, leading to a mismodeling that is particularly apparent at large values of the distribution of $\Delta=\abs{\ptmisscalo - \recoil} / \recoil$. This quantity tests the difference, relative to \recoil, between the particle-based \recoil and $\ptmisscalo$, which is \ptmiss measured only by the calorimeter systems and corrected for the presence of muons. To remove these events, $\Delta$ is required to be less than 0.5. 
Events are also vetoed if they contain an isolated photon passing loose identification criteria~\cite{CMS:2020uim} and with $\pt>25\GeV$ and $\abs{\eta}<2.5$. This requirement suppresses events from EW processes with an initial-state photon and from multijet production. 
Backgrounds from top-quark processes are reduced by vetoing any event containing b-tagged jets, defined as jets originating from the hadronization of b quarks using a combined secondary vertex discriminator with the medium working point~\cite{CMS:2017wtu}.
Further reduction of the QCD multijet background is achieved by requiring the azimuthal separation between each of the four leading jets and \recoilvec (\dphimin) to be greater than 0.5.
Additional selection requirements for each region on top of this baseline selection are described in the following. 

For the  \ptmiss plus jets region, backgrounds from \PW\ and \PZ bosons decaying via charged leptons are suppressed by vetoing events containing one or more isolated muon or electron passing a loose identification selection~\cite{CMS:2019ied,CMS:2020uim} with $\pt>10\GeV$ and $\abs{\eta}<2.5$ or a \tauh with $\pt>20\GeV$ and $\abs{\eta}<2.3$.

The dilepton region is selected by requiring exactly one pair of isolated muons or electrons having opposite electric charges, such that the invariant mass (\mll) of the pair is compatible with the \PZ boson mass ($71 < \mll < 111\GeV$). The muons and electrons are required to have $\pt>25$ and $ > 30\GeV$ respectively, and pass a medium identification selection~\cite{CMS:2019ied,CMS:2020uim}. For \dymm (\dyee) events, electron (muon) and \tauh vetoes are applied.

The \mujets region is selected by requiring exactly one well identified and isolated muon with $\pt>25\GeV$ and $\abs{\eta}<2.4$.
The corresponding \pt requirement for electrons in the \ejets region is $30\GeV$. The \tauhjets region is selected by requiring one $\tauh$ lepton passing a tight identification selection~\cite{CMS:2018jrd} with $\pt>40\GeV$ and $\abs{\eta}<2.3$. 
The transverse mass, defined as $\mT = \sqrt{\smash[b]{2\ptmiss\pt^{\ell}(1-\cos\Delta\phi)}}$, where $\Delta\phi$ is the azimuthal angle between the lepton and \ptmiss, is required to be in the range $30<\mT<125\GeV$. In addition, for the \ejets region, \ptmiss is required to be $>100\GeV$ to reduce the contribution from QCD multijet events. 

The largest background contribution to the \ptmiss plus jets region comes from \wjets events, accounting for 37\% of the events in this region. This includes events where the \PW boson decays via electrons (6.6\%), muons (9.6\%), hadronically decaying taus (13.1\%), and leptonically decaying taus (7.3\%).
This background is estimated using the control regions of \mujets and \ejets events that are independent of the signal region, but have similar kinematic requirements. These control regions select events where there is a single muon or electron in the final state, from both a \PW boson decaying directly to an electron or muon or indirectly via a leptonically decaying tau lepton. 

The estimation of this background proceeds by defining a transfer factor (\tw) as the ratio of the number of \wjets events in the signal region to that in the control region, as predicted by simulation. This \tw is determined as a function of \recoil and used to extrapolate from the event yields in data in the \mujets and \ejets control regions to obtain the number of expected \wjets events in the signal region. Systematic uncertainties in \tw associated with the muon and electron reconstruction, identification, and isolation requirements are included, and contribute at the 2\% level for muons and electrons. Uncertainties in \tw associated with the JES and JER, the \ptmiss trigger and electron trigger efficiencies, pileup, and b-tagged jet veto are also included. The jet energy scale uncertainty does not completely cancel in the transfer factor and has a residual effect of a few percent at low \recoil. Theoretical uncertainties in the corrections applied to account for NNLO QCD and NLO EW contributions, and from PDFs are also included; however, these mostly cancel in the ratio of \wjets events in the signal and control regions. The contributions from all these sources of uncertainties on \tw are included in Table~\ref{systs}

\begin{figure*}[htb!]
    \centering
    \includegraphics[width=1.0\textwidth]{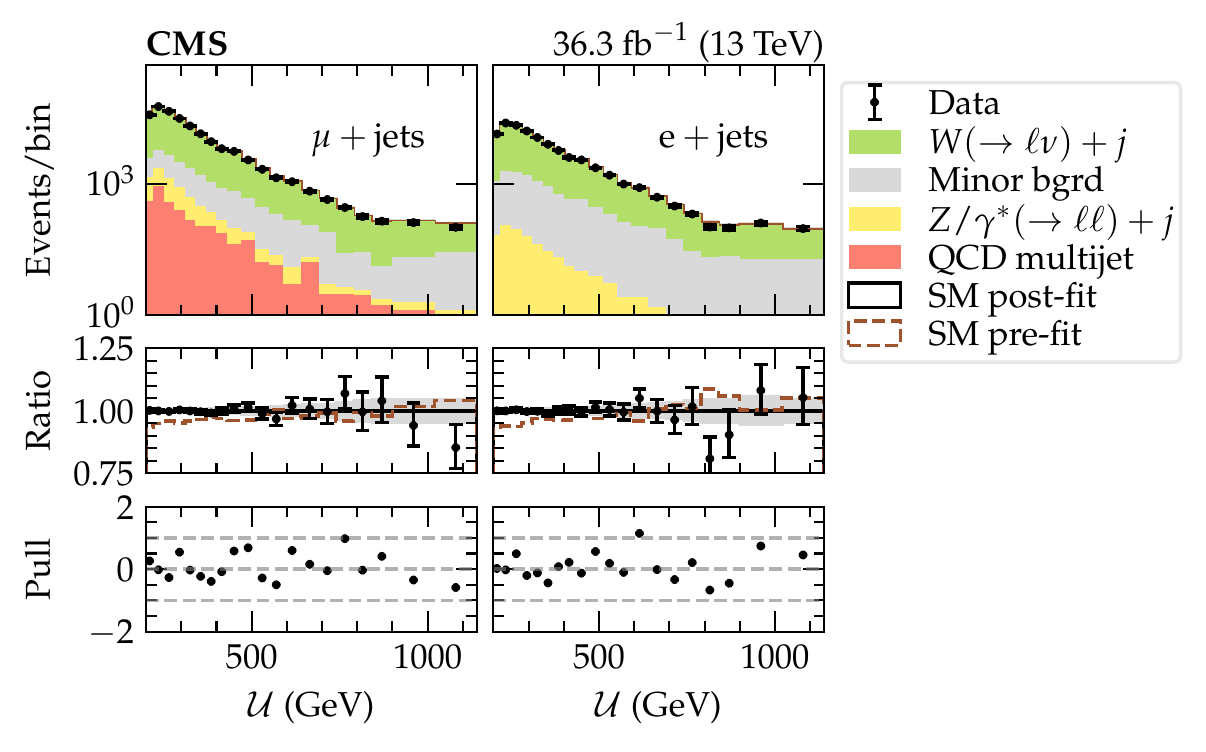}
    \caption{
    Distributions of \recoil in the regions of \mujets (left) and \ejets (right). Ratios are taken of the data with respect to both the SM post-fit and SM pre-fit results from a simultaneous likelihood fit performed to assess the consistency between the \mujets and \ejets channels. Pull is the difference between the data and the SM post-fit results, normalized by their uncertainties summed in quadrature. The `Minor' background includes the sum of all the contributions from processes such as diboson events, top pair production, and single top production.
    }
    \label{fig:tf_mue}
\end{figure*}

Studies are performed to validate the transfer factor method and to further test the extrapolation involved in predicting \wjets events in the signal region from the combined \mujets and \ejets regions. A simultaneous likelihood fit is performed to assess the consistency between the \mujets and \ejets channels by using the transfer factor method to predict the event yields in \ejets region from the \mujets region. The fit has two unconstrained parameters that are used to extract differences in the overall normalization and shape of the \recoil distribution with all systematic uncertainties included as nuisance parameters. The post-fit values for these parameters are $1.000\pm0.035$ and $-0.019\pm0.020$ for the normalization and shape parameters, respectively. The shape parameter is constructed as a dimensionless parameter that varies the shape of the \recoil distribution in one region linearly with respect to the other, effectively giving the value of the slope for the ratio of the \recoil distribution in the two regions. Values of 1.0 for the normalization parameter, and 0.0 for the shape parameter signify perfect consistency between the two regions. Hence, the post-fit values are in agreement within uncertainties with the expectation for consistent electron and muon channels. Figure~\ref{fig:tf_mue} shows the distributions of \recoil in the \mujets and \ejets regions before and after the likelihood fit.
The dominant uncertainties are from the electron trigger efficiency, followed by the muon and electron identification efficiencies.

\begin{figure*}[htb!]
    \centering
    \includegraphics[width=1.0\textwidth]{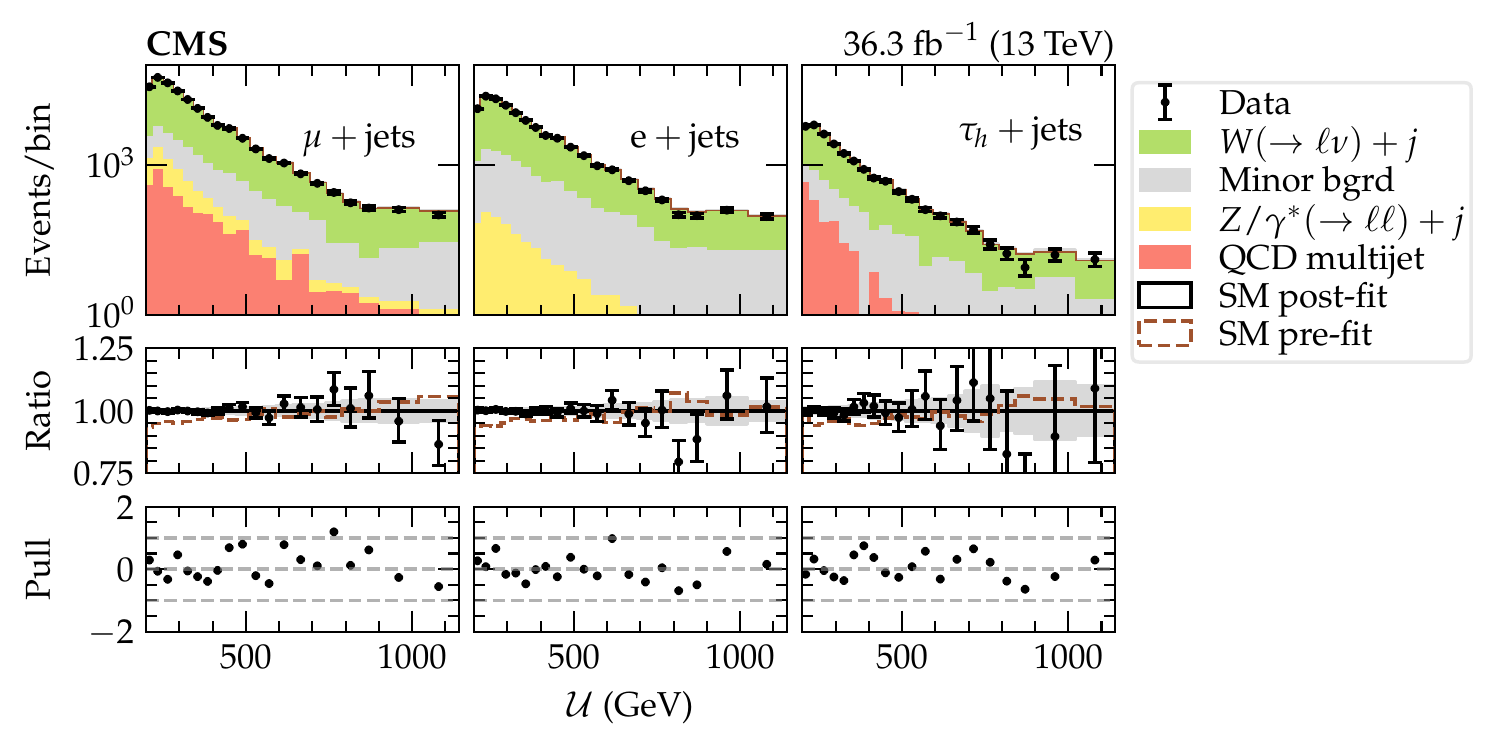}
    \caption{Distributions of \recoil in the regions of \mujets (left), \ejets (center), and \tauhjets (right). Ratios are taken of the data with respect to both the SM post-fit and SM pre-fit results from a likelihood fit performed to verify the consistency between the combined \mujets and \ejets region with the \tauhjets region. The pulls are the difference between the data and the SM post-fit results, normalized by their uncertainties summed in quadrature. The `Minor' background includes the sum of all the contributions from processes such as diboson events, top pair production, and single top production.}
    \label{fig:tf_all}
\end{figure*}

A similar likelihood fit is performed to verify the consistency between the combined \mujets and \ejets region with the \tauhjets region with two unconstrained parameters to extract differences in the normalization and shape of the \recoil distribution. Again, the post-fit values of these parameters show consistency within the uncertainties among all three regions, with values of $1.009\pm0.052$ and $0.001\pm0.044$ for the normalization and shape parameters, respectively. Figure~\ref{fig:tf_all} shows the distributions of \recoil in the \mujets, \ejets and \tauhjets regions before and after the likelihood fit. All systematic uncertainties are included as nuisance parameters in the fit, with the largest uncertainty coming from the \tauh selection efficiency and subdominant sources of uncertainty from the muon identification and electron trigger efficiencies. 

The contribution to the signal region from QCD multijet events is estimated from a control region that is selected by applying the same criteria as to the signal region but inverting the requirement on \dphimin. A jet from a QCD multijet event that is severely mismeasured results in significant \ptmiss and will have a direction closely aligned with \ptvecmiss. If there are one or more jets with $\dphimin < 0.5$, and if at least one of those jets has $\pt>40\GeV$, then the event falls into the QCD control region. Most events in the QCD control region have just one jet with $\dphimin < 0.5$ and $\pt>40\GeV$. A fit to the \pt distribution of this \ptmiss-aligned jet is used to extrapolate the event yields for QCD multijet events from the control region to the signal one.
An overall disagreement of 17\% is seen in the normalization in this QCD control region between data and simulation, attributed to the mismodelling of the QCD process in simulation. A likelihood fit is performed where this disagreement is parameterised as a linear fit to the \pt distribution of this \ptmiss-aligned jet in each \recoil bin and extrapolated to the signal region for $\pt<40\GeV$. This linear fit parameterisation significantly improves the agreement between data and simulation and is used to extract a normalization scale factor for the QCD multijet process in the signal region. This scale factor ranges from $1.11\pm0.08$ in the lowest \recoil bin to $2.44\pm0.99$ for the last \recoil bin. The total uncertainty in this normalization scale factor includes contributions from the jet energy scale and resolution, pileup, and lepton-related uncertainties. An additional systematic uncertainty is also assigned from the difference between a linear fit and a quadratic fit to extrapolate from the multijet-enriched control region to the signal region. The overall contribution of the uncertainty on the QCD multijet background on the final result is negligible.

Other backgrounds from top quark pair, single top quark, diboson, and photon+jets production are taken from simulation. Since most of these processes are generated at NLO, the simulation is considered to be a reliable estimation of their contributions to the various signal and control regions.  

The contributions of virtual photon exchange must be corrected to extract a value for the invisible width. This requires estimates of the contributions from \gstarll and its interference with \zll.
These are determined from \MGvATNLO where the pure \zll process and \gstarll is generated and the ratio of the cross sections of these processes with respect to \dyll, within the invariant mass region $71<\mll<111\GeV$, is used to determine the contribution from \gstarll and the interference term. 
Systematic uncertainties from PDFs, renormalization and factorization scales, and limited size of MC samples are included. The contribution from \gstarll is 1.5\% and is subtracted as a background to the \dyll process, whereas the interference is 0.6\% and is used to scale the cross section for \dyll to account for the interference term appearing in the \zll amplitude.

The main systematic uncertainties affecting the measurement of the \PZ invisible width are described in the following. 
Systematic uncertainties from the muon and electron identification efficiencies, obtained from Ref.~\cite{CMS:2019raw}, are among the dominant uncertainties, contributing 2.1 and 1.6\%, respectively. The uncertainty in the jet energy scale is the second-largest uncertainty, at 1.9\%. The statistical uncertainty in the electron identification efficiency and the uncertainty from pileup both contribute at the 1\% level, whereas the uncertainty from the $\tauh$ veto efficiency is 0.7\%. Experimental uncertainties from the \ptmiss and electron trigger efficiencies are at the 0.7\% level, whereas other experimental uncertainties from the jet energy resolution, QCD multijet background, and the b-tagged jet veto efficiency are negligible.  
Theoretical uncertainties from higher-order QCD and EW corrections applied to the \zjets and \wjets processes are estimated using the prescription of Ref.~\cite{Lindert:2017olm} and implemented as five parameters. The final impact on the result is 0.5\%. Uncertainties in PDFs mostly cancel, with a residual contribution of 0.2\%. 
The combined statistical and systematic uncertainty in the measurement is 3.2\%, with the statistical uncertainty contributing 0.6\%. The main sources of systematic uncertainty and their relative contribution (in \%) are shown in Table~\ref{systs}.

\begin{table*}[htb!]
    \centering
    \topcaption{Relative uncertainties (in \%) on the final measurement from different sources.}
    \begin{tabular}{ll}
        \hline
        Source of systematic uncertainty & Uncertainty (\%)\\
        \hline
        Muon identification efficiency (syst.) & 2.1 \\
        Jet energy scale & 1.8--1.9 \\
        Electron identification efficiency (syst.) & 1.6\\
        Electron identification efficiency (stat.) & 1.0\\
        Pileup & 0.9--1.0 \\
        Electron trigger efficiency & 0.7\\
        $\tau_{h}$ veto efficiency & 0.6--0.7 \\
        \ptmiss trigger efficiency (jets plus \ptmiss region) & 0.7 \\
        \ptmiss trigger efficiency (\dymm region) & 0.6 \\
        Boson \pt dependence of QCD corrections  & 0.5\\
        Jet energy resolution & 0.3--0.5\\
        \ptmiss trigger efficiency (\mujets region) & 0.4\\
        Muon identification efficiency (stat.) & 0.3 \\
        Electron reconstruction efficiency (syst.) & 0.3\\
        Boson \pt dependence of EW corrections  & 0.3\\
        PDFs & 0.2\\
        Renormalization/factorization scale & 0.2 \\
        Electron reconstruction efficiency (stat.) & 0.2\\ [\cmsTabSkip]
        Overall & 3.2 \\
        \hline
    \end{tabular}
    \label{systs}
\end{table*}

\begin{figure*}[htb!]
    \centering
    \includegraphics[width=1.0\textwidth]{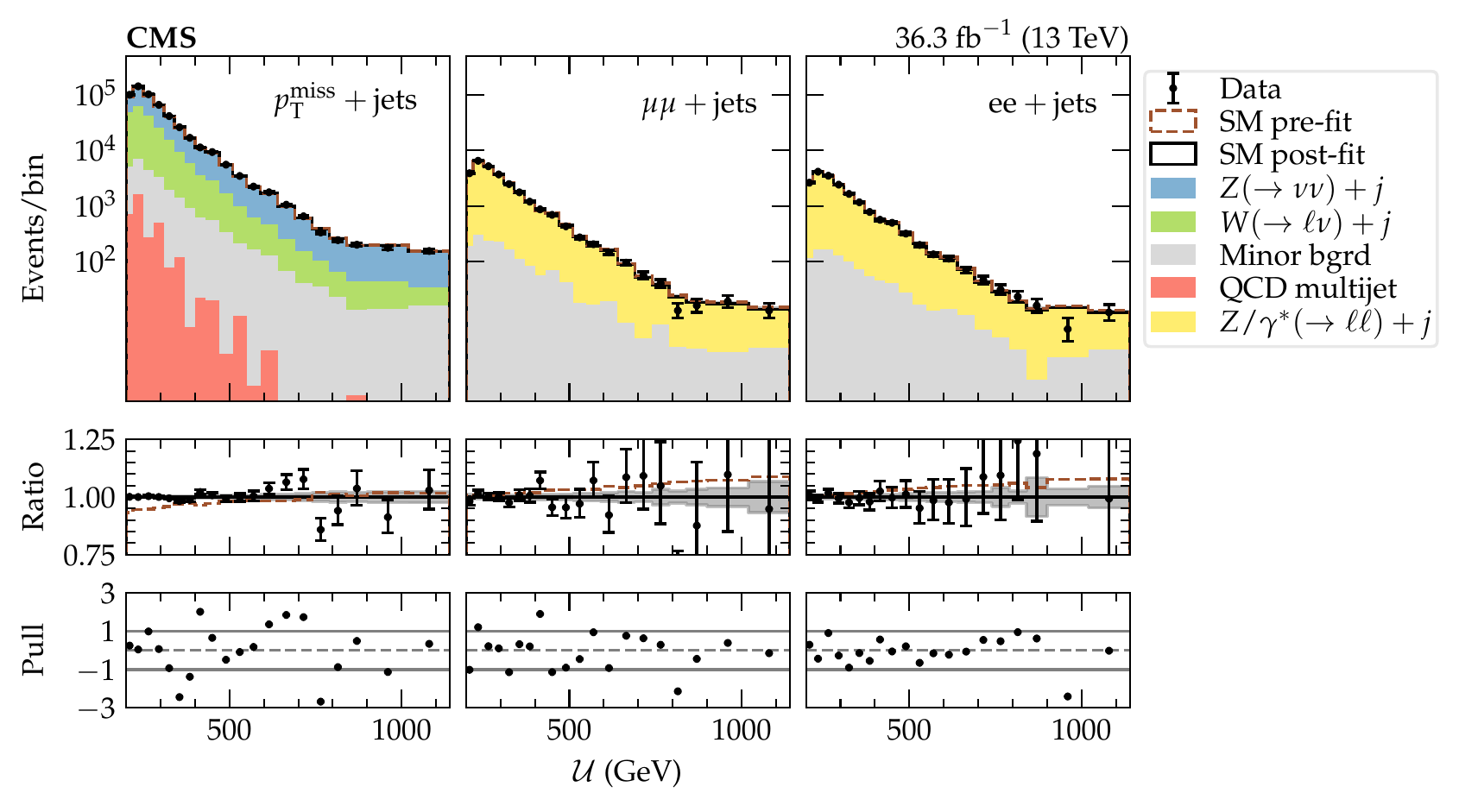}
    \caption{Distribution of \recoil for the \ptmissjets (left), \dymm (center), and \dyee (right) regions. Ratios are taken of the data with respect to both the SM post-fit and SM pre-fit results. The pulls are the difference between the data and the SM post-fit results, normalized by their uncertainties summed in quadrature. The `Minor' background includes the sum of all the contributions from processes such as diboson events, top pair production, single top production that are determined from simulation. }
    \label{fig:postfit}
\end{figure*}

The invisible width of the \PZ boson is extracted from a simultaneous likelihood fit to the \recoil distributions in the \ptmissjets, \dymmjets, \dyeejets, \mujets, and \ejets regions. The \recoil distribution is divided into 20 bins to exploit the signal-to-background ratios and the dependence of systematic uncertainties on \recoil. The bin widths are chosen to be roughly equivalent to the \recoil resolution and with a maximum statistical uncertainty of 30\% for the \dymmjets and \dyeejets regions. 
The transfer factor estimating the \wjets background is implemented as a global unconstrained parameter scaling the \wjets process in the \ptmissjets and $\ell+$jets regions and an additional free parameter is used to scale the normalisation of both the \zll and \zvv processes. Systematic uncertainties are treated as nuisance parameters, modeled with Gaussian constraints. The post-fit distributions from the simultaneous fit are shown in Fig.~\ref{fig:postfit}. The total number of events observed in data and from the the post-fit simulation results for the \ptmissjets, \dymmjets, and \dyeejets regions are shown in Table~\ref{summary}.

The parameter of interest, \rinv, is defined as \widthinv/\widthinvMC, using the ratio of Eq.~\ref{eq:zinv-width} with its expectation from simulation and assuming that the value of $\Gamma(\zll)$ in data is the same as simulation. 
The value of \rinv is measured to be $1.049\pm0.006\stat\pm0.032\syst$. This is translated into a measurement of the width using the \zvv partial width from simulation (\widthinvMC), which is calculated from the branching fraction of \zvv in \MGvATNLO~\cite{Alwall:2014hca} for an invariant mass range of 71--111\GeV and the total \PZ boson width of 2.4988\GeV. The value of \widthinvMC is 498.7\MeV and the \zll partial width to one charged lepton flavor is 84.2\MeV.

\begin{table*}[htb!]
    \centering
    \topcaption{SM post-fit predictions for the \ptmissjets, \dymm, and \dyee regions and the total number of events measured in data.}
    \begin{tabular}{lccc}
        \hline
        & \ptmissjets & \dymm & \dyee \\
        \hline
        \zvvjets & 310000 & \NA & \NA  \\
        \dyll &  2680    & 25900 & 17300 \\
        \wlnujets & 195000 &\NA  & \NA \\
        QCD multijet & 3360 & \NA  & \NA \\
        Minor & 25600 & 1720 & 1090 \\ [\cmsTabSkip]
        Total SM &  537000 & 27600 & 18400       \\ [\cmsTabSkip]
        Data & 537326 & 27631  & 18326 \\
        \hline
    \end{tabular}
    \label{summary}
\end{table*}

The measured \PZ boson invisible width value (Eq.~\ref{eq:zinv-width}) is obtained by multiplying \rinv by \widthinvMC:
\begin{equation}
    \centering
    \widthinv = 523\pm 3\stat \pm 16\syst \MeV.
\end{equation}

\begin{figure}[htb!]
    \centering
    \includegraphics[width=\cmsFigWidth]{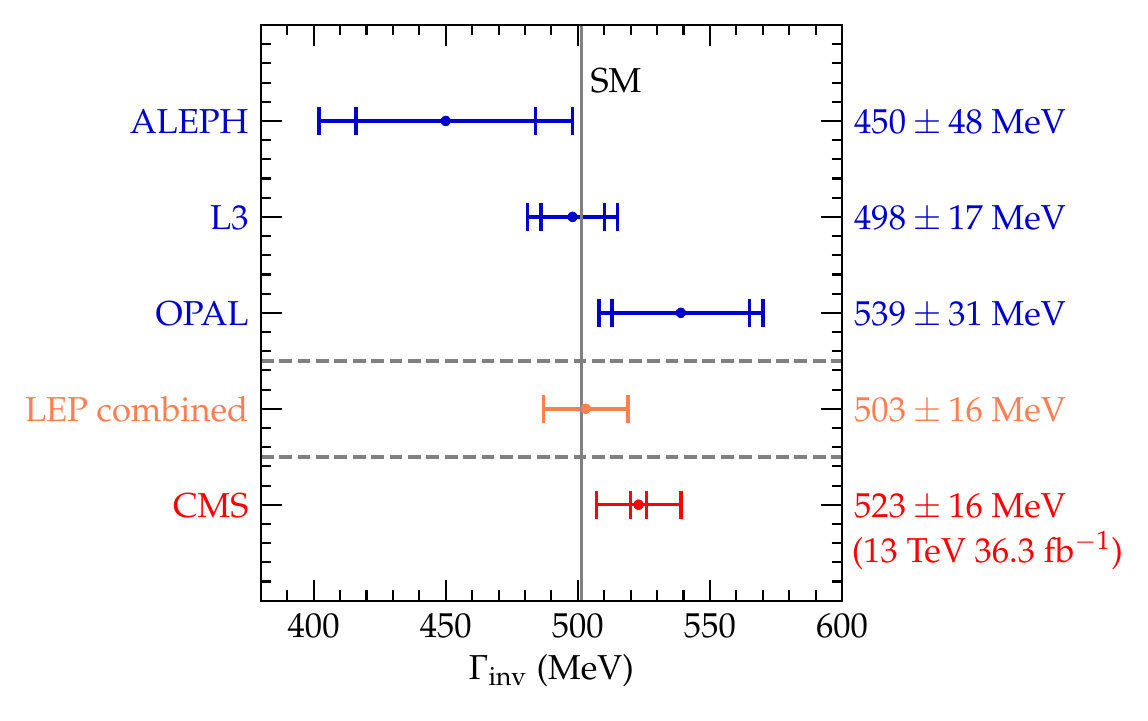}
    \caption{Direct measurements of the Z invisible width by the LEP experiments and the result from the CMS experiment presented here. Also shown is the prediction from the SM. The inner (outer) error bars show the statistical (total) uncertainty.}
    \label{fig:summary}
\end{figure}

A comparison with the direct measurements from the LEP experiments ALEPH, OPAL, and L3, as well as with the LEP combined measurement is shown in Fig.~\ref{fig:summary}. Also shown is the standard model prediction for $\Gamma(\zvv)$ of $501.44\pm0.04\MeV$~\cite{ALEPH:2005ab,10.1093/ptep/ptaa104}. With an uncertainty of 16\MeV, this measurement of the \PZ boson invisible width at CMS is comparable in precision to the LEP combined result and is compatible with the value expected in the SM. 

In summary, this Letter presents the first precise measurement of the invisible width (\widthinv) of the \PZ boson at a hadron collider. The measurement is performed using proton-proton collisions at $\sqrt{s} = 13\TeV$ and data corresponding to an integrated luminosity of 36.3\fbinv recorded with the CMS detector, yielding a result of $\widthinv=523\pm 3\stat \pm 16\syst\MeV$. It represents the single most precise direct measurement of \widthinv to date and is competitive with the combined direct measurement from LEP and compatible with the value expected in the standard model.

\begin{acknowledgments}
We congratulate our colleagues in the CERN accelerator departments for the excellent performance of the LHC and thank the technical and administrative staffs at CERN and at other CMS institutes for their contributions to the success of the CMS effort. In addition, we gratefully acknowledge the computing centers and personnel of the Worldwide LHC Computing Grid and other centers for delivering so effectively the computing infrastructure essential to our analyses. Finally, we acknowledge the enduring support for the construction and operation of the LHC, the CMS detector, and the supporting computing infrastructure provided by the following funding agencies: BMBWF and FWF (Austria); FNRS and FWO (Belgium); CNPq, CAPES, FAPERJ, FAPERGS, and FAPESP (Brazil); MES and BNSF (Bulgaria); CERN; CAS, MoST, and NSFC (China); MINCIENCIAS (Colombia); MSES and CSF (Croatia); RIF (Cyprus); SENESCYT (Ecuador); MoER, ERC PUT and ERDF (Estonia); Academy of Finland, MEC, and HIP (Finland); CEA and CNRS/IN2P3 (France); BMBF, DFG, and HGF (Germany); GSRI (Greece); NKFIH (Hungary); DAE and DST (India); IPM (Iran); SFI (Ireland); INFN (Italy); MSIP and NRF (Republic of Korea); MES (Latvia); LAS (Lithuania); MOE and UM (Malaysia); BUAP, CINVESTAV, CONACYT, LNS, SEP, and UASLP-FAI (Mexico); MOS (Montenegro); MBIE (New Zealand); PAEC (Pakistan); MES and NSC (Poland); FCT (Portugal); MESTD (Serbia); MCIN/AEI and PCTI (Spain); MOSTR (Sri Lanka); Swiss Funding Agencies (Switzerland); MST (Taipei); MHESI and NSTDA (Thailand); TUBITAK and TENMAK (Turkey); NASU (Ukraine); STFC (United Kingdom); DOE and NSF (USA).

\hyphenation{Rachada-pisek} Individuals have received support from the Marie-Curie program and the European Research Council and Horizon 2020 Grant, contract Nos.\ 675440, 724704, 752730, 758316, 765710, 824093, 884104, and COST Action CA16108 (European Union); the Leventis Foundation; the Alfred P.\ Sloan Foundation; the Alexander von Humboldt Foundation; the Belgian Federal Science Policy Office; the Fonds pour la Formation \`a la Recherche dans l'Industrie et dans l'Agriculture (FRIA-Belgium); the Agentschap voor Innovatie door Wetenschap en Technologie (IWT-Belgium); the F.R.S.-FNRS and FWO (Belgium) under the ``Excellence of Science -- EOS" -- be.h project n.\ 30820817; the Beijing Municipal Science \& Technology Commission, No. Z191100007219010; the Ministry of Education, Youth and Sports (MEYS) of the Czech Republic; the Hellenic Foundation for Research and Innovation (HFRI), Project Number 2288 (Greece); the Deutsche Forschungsgemeinschaft (DFG), under Germany's Excellence Strategy -- EXC 2121 ``Quantum Universe" -- 390833306, and under project number 400140256 - GRK2497; the Hungarian Academy of Sciences, the New National Excellence Program - \'UNKP, the NKFIH research grants K 124845, K 124850, K 128713, K 128786, K 129058, K 131991, K 133046, K 138136, K 143460, K 143477, 2020-2.2.1-ED-2021-00181, and TKP2021-NKTA-64 (Hungary); the Council of Science and Industrial Research, India; the Latvian Council of Science; the Ministry of Education and Science, project no. 2022/WK/14, and the National Science Center, contracts Opus 2021/41/B/ST2/01369 and 2021/43/B/ST2/01552 (Poland); the Funda\c{c}\~ao para a Ci\^encia e a Tecnologia, grant CEECIND/01334/2018 (Portugal); the National Priorities Research Program by Qatar National Research Fund; MCIN/AEI/10.13039/501100011033, ERDF ``a way of making Europe", and the Programa Estatal de Fomento de la Investigaci{\'o}n Cient{\'i}fica y T{\'e}cnica de Excelencia Mar\'{\i}a de Maeztu, grant MDM-2017-0765 and Programa Severo Ochoa del Principado de Asturias (Spain); the Chulalongkorn Academic into Its 2nd Century Project Advancement Project, and the National Science, Research and Innovation Fund via the Program Management Unit for Human Resources \& Institutional Development, Research and Innovation, grant B05F650021 (Thailand); the Kavli Foundation; the Nvidia Corporation; the SuperMicro Corporation; the Welch Foundation, contract C-1845; and the Weston Havens Foundation (USA).    
\end{acknowledgments}

\bibliography{auto_generated}

\cleardoublepage \appendix\section{The CMS Collaboration \label{app:collab}}\begin{sloppypar}\hyphenpenalty=5000\widowpenalty=500\clubpenalty=5000
\cmsinstitute{Yerevan Physics Institute, Yerevan, Armenia}
{\tolerance=6000
A.~Tumasyan\cmsAuthorMark{1}\cmsorcid{0009-0000-0684-6742}
\par}
\cmsinstitute{Institut f\"{u}r Hochenergiephysik, Vienna, Austria}
{\tolerance=6000
W.~Adam\cmsorcid{0000-0001-9099-4341}, J.W.~Andrejkovic, T.~Bergauer\cmsorcid{0000-0002-5786-0293}, S.~Chatterjee\cmsorcid{0000-0003-2660-0349}, K.~Damanakis\cmsorcid{0000-0001-5389-2872}, M.~Dragicevic\cmsorcid{0000-0003-1967-6783}, A.~Escalante~Del~Valle\cmsorcid{0000-0002-9702-6359}, R.~Fr\"{u}hwirth\cmsAuthorMark{2}\cmsorcid{0000-0002-0054-3369}, M.~Jeitler\cmsAuthorMark{2}\cmsorcid{0000-0002-5141-9560}, N.~Krammer\cmsorcid{0000-0002-0548-0985}, L.~Lechner\cmsorcid{0000-0002-3065-1141}, D.~Liko\cmsorcid{0000-0002-3380-473X}, I.~Mikulec\cmsorcid{0000-0003-0385-2746}, P.~Paulitsch, F.M.~Pitters, J.~Schieck\cmsAuthorMark{2}\cmsorcid{0000-0002-1058-8093}, R.~Sch\"{o}fbeck\cmsorcid{0000-0002-2332-8784}, D.~Schwarz\cmsorcid{0000-0002-3821-7331}, S.~Templ\cmsorcid{0000-0003-3137-5692}, W.~Waltenberger\cmsorcid{0000-0002-6215-7228}, C.-E.~Wulz\cmsAuthorMark{2}\cmsorcid{0000-0001-9226-5812}
% ...
\par}
\vskip\cmsinstskip
\dag:~Deceased\\
$^{1}$Also at Yerevan State University, Yerevan, Armenia\\
$^{2}$Also at TU Wien, Vienna, Austria\\
% ...
\end{sloppypar}
\end{document}